\documentclass{osa-article}
\journal{osajournal}
\articletype{Research Article}

\newcommand{\name}{ONIX\xspace}

\usepackage{amsmath,bm}
\usepackage{xspace}

\newcommand{\mymath}[2]{\newcommand{#1}{\TextOrMath{$#2$\xspace}{#2}}}

\mymath{\contrast}{c}
\mymath{\viewIndex}{m}
\mymath{\ior}{n}
\mymath{\worldSpaceCoord}{\mathbf{x}}
\mymath{\viewSpaceCoord}{\mathbf{y}}
\mymath{\worldToViewSpace}{\mathrm{\mathbf G}}
\mymath{\encoder}{\mathrm{\mathbf E}}
\mymath{\mlp}{\mathrm{\mathbf F}}
\mymath{\prior}{\mathbf{p}}
\mymath{\allKnownView}{V}
\mymath{\priorView}{v}
\mymath{\ray}{\mathbf{r}}
\mymath{\allRays}{R}
\mymath{\rayOrigin}{\mathbf{o}}
\mymath{\rayDirection}{\mathbf{d}}
\mymath{\coordinateX}{x}
\mymath{\coordinateY}{y}
\mymath{\coordinateZ}{z}
\mymath{\thickness}{t}
\mymath{\posionalEncoder}{\gamma}
\mymath{\genintensity}{\widehat{c}}
\mymath{\imag}{i}
\mymath{\realIOR}{\delta}
\mymath{\imagIOR}{\beta}
\mymath{\distBetweenAdjacentSample}{\Delta}
\mymath{\degree}{^{\circ}}

\def\resultPath{figures/Results/}
\def\figurePath{figures/}
\DeclareGraphicsExtensions{.png,.jpg,.jpeg,.pdf,.ai,.psd}
\usepackage[nolist]{acronym}
\begin{acronym}
\acro{DL}{Deep learning}
\acro{ML}{Machine learning}
\acro{DSSIM}{Dissimilarity Structure Similarity Index Metric}
\acro{FRCM}{Fourier Ring Correlation Metric}
\acro{FRC}{Fourier Ring Correlation}
\acro{CNN}{Convolutional Neural Network}
\acro{MLP}{Multilayer Perceptron}
\acro{XFEL}{X-ray Free-electron Laser}
\acro{MSE}{Mean-square-error}
\acro{SART}{Simultaneous Algebraic Reconstruction Technique}
\end{acronym}

\begin{document}

\title{\name: an X-ray deep-learning tool for 3D reconstructions from sparse views} 

\author{Yuhe Zhang,\authormark{1,*} Zisheng Yao,\authormark{1}, Tobias Ritschel, \authormark{2}, and Pablo Villanueva-Perez\authormark{1}}

\address{\authormark{1}Synchrotron Radiation Research and NanoLund, Lund University, Box 118, 221 00, Lund, Sweden\\
\authormark{2}University College London, WC1E 6BT London, UK\\}

\email{\authormark{*}yuhe.zhang@sljus.lu.se}


\begin{abstract}
Three-dimensional (3D) X-ray imaging techniques like tomography and confocal microscopy are crucial for academic and industrial applications. 
These approaches access 3D information by scanning the sample with respect to the X-ray source. 
However, the scanning process limits the temporal resolution when studying dynamics and is not feasible for some applications, such as surgical guidance in medical applications.
Alternatives to obtaining 3D information when scanning is not possible are X-ray stereoscopy and multi-projection imaging.
However, these approaches suffer from limited volumetric information as they only acquire a small number of views or projections compared to traditional 3D scanning techniques.
Here, we present \name (Optimized Neural Implicit X-ray imaging), a deep-learning algorithm capable of retrieving 3D objects with arbitrary large resolution from only a set of sparse projections.
\name, although it does not have access to any volumetric information, outperforms current 3D reconstruction approaches because it includes the physics of image formation with X-rays, and it generalizes across different experiments over similar samples to overcome the limited volumetric information provided by sparse views.
We demonstrate the capabilities of \name compared to state-of-the-art tomographic reconstruction algorithms by applying it to simulated and experimental datasets, where a maximum of eight projections are acquired.
We anticipate that \name will become a crucial tool for the X-ray community by i) enabling the study of fast dynamics not possible today when implemented together with X-ray multi-projection imaging, and ii) enhancing the volumetric information and capabilities of X-ray stereoscopic imaging in medical applications.
\end{abstract}

\section{Introduction}
As discovered by Roentgen in 1895, the high penetration power of X-rays makes them an excellent probe to access volumetric and internal information for a plethora of materials and systems.
This property led to the development of three-dimensional (3D) techniques like tomography~\cite{Cormack1964CT,Hounsfield1973CT} and confocal microscopy~\cite{Davidovits1969Confocal,Ale2012ConfocalFluorescence}, which have become standard probes to access structural information in a non-destructive manner for many and diverse disciplines~\cite{Salvo2003CTinMS,Galban2012CTMedicine,Ale2012CTanimals,Withers2021CTreview}, e.g., medicine, biology, chemistry, physics, cultural heritage, materials science, geoscience, and industrial processes.  
These state-of-the-art 3D X-ray imaging techniques access volumetric information by scanning the sample over different exposures, which is a time-consuming process.
Thus, the scanning process may hamper the applicability of these 3D imaging approaches to i) 3D temporally-resolved studies~\cite{Garcia-Moreno2019FastCT,Ziesche2020FastCTBatteries}, where the scanning process takes longer than the studied dynamics or the scanning may induce forces that alter the studied dynamics, ii) experiments where the setup or configuration makes the scanning process difficult or impossible~\cite{Evans2003StereoDynamic,Verellen2003StereoMedical}, e.g., in medical surgeries and guidance approaches, and iii)  dose-limited experiments~\cite{Wang2006LowdoseCT,Zanette2012CTLow-dose},  for example medical applications or radiosensitive samples.

An alternative to avoid the scanning process is to use techniques that illuminate the sample simultaneously from different views, like X-ray stereographic methods~\cite{thomson1896stereoscopic}, from only two views, or X-ray multi-projection methods, using multiple views~\cite{Howells1994XMPI,villanueva2018hard}. 
These methods have been demonstrated and applied using diverse sources for different purposes.
For example, they have been used together with laboratory or conventional X-ray sources~\cite{Moseley1962StereoXrayFirstDevice} for positioning and guidance of medical surgeries or to capture fast dynamics~\cite{Evans2003StereoDynamic,Verellen2003StereoMedical}.
Lately, their application in conjunction with X-ray high-brilliance sources, i.e., X-ray sources with a high-coherent flux, such as diffraction-limited storage rings~\cite{Eriksson2014DLSR} and \acp{XFEL}~\cite{McNeil2010XFELreview,Emma2010XFEL_operation,SACLA2012Huang,Kang2017PALXFEL,Prat2020SwissFEL} has been proposed and demonstrated~\cite{villanueva2018hard,Duarte2019XStereo,Sowa2020Plenoptic,Voegeli2020XMPI,Goldberger20203DPtycho}.
When applied at such facilities, these non-scanning techniques open up the possibility to access temporal and spatial resolutions not possible before.
Moreover, they can obtain, for the first time, 3D information from single intense pulses from \acp{XFEL} for stochastic and non-reproducible samples when operated in diffraction-before-destruction mode~\cite{Neutze2000SPI,Chapman2006DiffractionBeforeDestructionExp}, i.e., the sample is destroyed after being illuminated with a single \ac{XFEL} pulse.
Although stereographic and multi-projection approaches can overcome the aforementioned limitations of state-of-the-art 3D X-ray imaging approaches, they can only access limited volumetric via a sparse number of views or projections.

A solution to overcome the limited 3D information obtained by stereo or multi-projection imaging techniques is to add prior knowledge about our sample and system.
Many frameworks have tried to address this issue, like compressed-sensing~\cite{Candes2006CS,Donoho2006CS}, and \ac{DL} approaches~\cite{LeCun2015DL}.
Specifically, the latter has demonstrated its ability to generate complex 3D scenes or volumes from sparse views in computer vision~\cite{kar2017learning}.
Some of the most successful \ac{DL} approaches to this problem rely on convolutional neural networks and representation of the 3D volume using regular grids \cite{wu2016learning} or irregular point clouds \cite{qi2017pointnet}. 
However, these representations and approaches have a large memory consumption limiting their applicability to small volumes, and so are not suitable for the aforementioned applications.
Recently, neural implicit 3D representations~\cite{chen2019learning,park2019deepsdf} have been developed to overcome this limitation by modeling 3D volumes at arbitrary resolution through a continuous space function.
Although these implicit 3D representations have been a paradigm shift in computer vision and enable 3D reconstructions with only two-dimensional (2D) supervision~\cite{mildenhall2020nerf,yu2021pixelnerf,henzler2021unsupervisedvideos}, they have not made an impact on the X-ray community yet.

\begin{figure*}[htbp!]
\centering\includegraphics*[width = \linewidth]{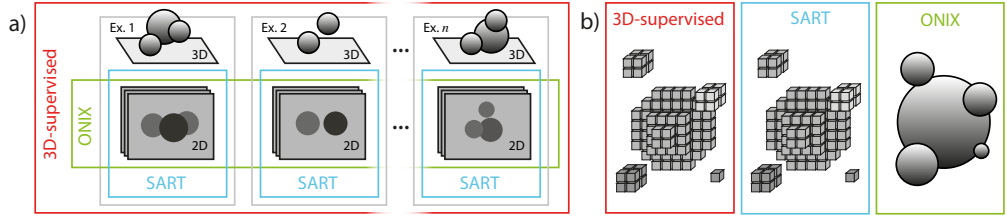}
\caption{
Comparison and two main advantages of our approach over previous work.
First, a) shows three 3D experiments with their corresponding 2D projections and how different approaches access this information.
 \ac{SART} (blue) and other conventional 3D methods only access 2D projections from a single 3D experiment at a time.
Typical \ac{DL} methods access all the projections from different 3D experiments but require 3D supervision (red), i.e., access the 3D volume itself.
\name (green) access all the projections from all 3D experiments, but without requiring 3D supervision at any point.
b) The second unique point of our approach is that it does not require any grid but works directly on implicit functions.
\label{fig:Concept}}
\end{figure*}

This paper presents \name (Optimized Neural Implicit X-ray imaging) an implicit 3D \ac{DL} approach for X-ray imaging that has the potential to retrieve arbitrary large resolution over 3D volumes from only a set of sparse 2D projections.
\name uses only one or a small set of 2D radiographs or projections to achieve 3D volume reconstructions extending the capabilities of current implicit representations by:
i) including an accurate physical description of the interaction of X-rays with matter via the projection approximation to impose self-consistency between the 3D reconstructed volume and the recorded radiographs, and ii) using a convolutional layer that enables transferring knowledge between similar X-ray multi-projection experiments to enhance the 3D reconstructions from limited views.
The interaction of X-rays with matter includes not only the attenuation term but also the phase term making \name compatible not only with standard attenuation radiographs but also with coherent~\cite{Chapman2010Coherent} and phase-sensitive techniques~\cite{Bravin2013PhaseContrast}.
We demonstrate the capabilities of \name by exploiting experimental and simulated data using up to a maximum of eight views in realistic configurations for X-ray multi-projection imaging.
The obtained results are compared to two grid approaches i) \ac{SART} a traditional iterative reconstruction approach from sparse views~\cite{Andersen1984SART}, and ii) a state-of-the-art 3D supervised machine learning approach~\cite{Liu2020TomoGAN} (see subsection~\ref{subsec:supervised}).
Figure~\ref{fig:Concept} depicts a schematic comparison between the three approaches that highlights the unique advantages of \name in terms of the inputs and outputs.
From this comparison, we conclude that \name will enable 3D reconstructions from sparse views at a quality not possible today without any 3D supervision.
Thus, it will become a tool that will expand the capabilities of current stereoscopic imaging techniques for medical applications and time-resolved experiments with laboratory sources, \acp{XFEL}, and diffraction-limited storage rings.

This paper is structured as follows: 
First, we describe the \name approach and how the physics of the X-ray interaction with matter is accounted for. 
We also introduce a supervised 3D \ac{DL} approach for comparison purposes.
Second, we validate \name with synthetic and experimental data using up to a maximum of eight projections, and we compare its performance to state-of-the-art 3D reconstruction algorithms and 3D supervised machine learning.
Finally, we discuss the results and future application of \name to enhance and enable 3D reconstructions from sparse projections coming from current X-ray stereoscopic and multi-projection approaches. 

\section{Methods}

\subsection{X-ray propagation and interaction with matter\label{subsec:interaction}}
The interaction between matter and electromagnetic radiation is usually described via the refractive index (\ior).
Unlike visible light, the refractive index for X-rays is very close to unity, and is usually expressed as a complex refractive index:
\begin{equation}
    \ior = 
    1-
    \realIOR + \imag \imagIOR~,
    \label{Eq: refractive_index}
\end{equation}
where \realIOR and \imagIOR are positive and real quantities that denote the phase-shifting and the absorptive part of the X-ray refractive index.
Due to the high penetration power and weak interaction of X-rays with matter, the values of \realIOR and \imagIOR are small.
This enables the projection approximation~\cite{Paganin2006CoherentX-ray}, which neglects diffraction within the sample volume. 
Under this approximation, the transmission of X-rays through a sample can be approximated by the exit wave immediately after the object:
\begin{equation}
    \psi_{\coordinateZ_{\rm{exit}}} = \psi_{\coordinateZ_0} \exp
    \left(
    -\imag k\int_{\coordinateZ=\coordinateZ_0}^{\coordinateZ=\coordinateZ_{\rm{exit}}}\big(\realIOR[\coordinateX,\coordinateY;\coordinateZ] - \imag \imagIOR[\coordinateX,\coordinateY;\coordinateZ]\big)
    \mathrm d\coordinateZ
    \right)~,
    \label{Eq: projection_appoximation}
\end{equation}
where $\psi_{z_0}$ is the incoming wave on the sample, $k=2\pi/\lambda$ is the wavenumber which is inversely proportional to the wavelength ($\lambda$), \realIOR and \imagIOR are a function of the transverse coordinates \coordinateX and \coordinateY orthogonal to the propagation direction \coordinateZ.

In the context of this work, we assume the projection approximation to describe the interaction of X-rays with matter. 
This approximation is used to transfer information between the 3D reconstructed volumes and the 2D recorded projections. 
It therefore enforces consistency between the 3D and 2D domains.

\subsection{Simulated data} \label{subsec:SimulatedData}

To demonstrate the capabilities and feasibility of \name, we generated a synthetic 3D dataset mimicking tomographic experiments and their measured projections. 
Each of the 3D objects consists of a collection of five to ten ellipsoids randomly positioned inside a cylinder. 
In our simulations, the ellipsoids were void, while the cylinder was made out of pure aluminum.
We assumed the index of refraction of aluminum at 18 keV, where $\realIOR = 1.6741 \times 10^{-7}$ and $\imagIOR = 6.4088 \times 10^{-9}$.
The void ellipsoids were simulated assuming $\ior=1$. 
The 3D volumes had $256\times256\times256$ voxels of 3.2~$\mu$m.
Each of the volumes contained the aluminum cylinder with a diameter of 100~voxels and a height of 256~voxels.
Inside the cylinder, the ellipsoids were positioned with semi-axes randomly generated between 20 to 80~voxels.
This synthetic dataset contains a total of 1000 of the previously described 3D objects.
For each object, we simulated the absorption and phase contrast projections from eight equally-spaced angles between 0$\degree$ and 140$\degree$. 
These projections were used as the input for the training of the \ac{DL} methods.

\subsection{Experimental data: metallic foams} \label{subsec:ExperimentalData}
We also validated the performance of \name using experimental data coming from X-ray tomography or 3D imaging on metallic foams. 
The X-ray tomographic experiments of metallic foams~\cite{Garcia-Moreno2019MetallicFoams} made out of thixo, a well-established compound for metallic foams, were performed at the TOMCAT beamline of the Swiss Light Source~\cite{Stampanoni2006TOMCAT}.
Each of these acquisitions used the polychromatic spectrum provided by the TOMCAT's bending magnet source.
This means that each radiograph only contains attenuation contrast as the coherence was not sufficient to access the phase information.
The radiographs were acquired with the GigaFRoST camera system that enabled 3D time-resolved acquisitions of the foaming process~\cite{Mokso2017Gigafrost}.
The camera system provided an effective pixel size of $2.75~\mu$m, and the radiographs had dimensions of $180\times960$ pixels.
To speed up the training, we downsized the projections into 180 $\times$ 256 frames, reducing the number of voxels.
A total of 168 tomographic experiments were used to train our \ac{DL} approaches.
From each of these experiments, we retrieved a 3D volume from a total of 96 radiographs with projection angles homogeneously distributed between 0 and 180~degrees.
Given the limited number of projections (below the Crowther criterion), we used the \ac{SART}~\cite{Andersen1984SART} to retrieve the 3D volumes.
\ac{SART} provides high-quality tomographic reconstructions from limited views using an iterative approach based on linear algebra methods. 
To train our \ac{DL} methods, we selected eight radiographs per volume with projection angles equally spaced between 0$\degree$ and 131$\degree$, from all the acquired projections.

\subsection{3D unsupervised learning: \name}
This subsection describes \name, our  \ac{DL} approach capable of reconstructing 3D volumes from only a set of sparse projections without accessing 3D information and with theoretically infinite resolution. 
First, we discuss the main features of \name. 
Then, we discuss the architecture and implementation details of \name and how it includes the physics of X-ray imaging and can learn across different samples and experiments to overcome the limitations of state-of-the-art 3D reconstruction approaches like \ac{SART}.
Finally, we describe the optimization process and specific parameters used for the aforementioned datasets.

Most 3D X-ray image reconstruction methods work on discrete representations, defined by grids of pixels and voxels.
This limits the achievable resolution, tends to be computationally demanding, and consumes large amounts of memory.
Though we are used to representing spatial information as grids, the physical world is primarily continuous, especially on the microscale.
Thus, one may consider using continuous and differentiable functions instead of discrete representations to describe nature.
This functional representation is called implicit representation \cite{xie2021neuralfield}.
Although implicit representations have an excellent potential for 3D X-ray imaging approaches, they have not been used, as finding a function that approximates a 3D sample of interest is a cumbersome task.
The advent of \ac{DL} methods, such as deep neural networks, has opened new opportunities to address this problem by exploiting their potential as universal approximators.
To approximate a 3D sample by a \ac{DL} approach, we need to optimize (train) its parameters.
This optimization is done by minimizing a cost function, which is usually called the "loss function" or "objective" of the neural network.
By finding a proper neural network and optimization approach, we can use \ac{DL} methods to infinitely approximate the shape or distribution of an object to overcome the aforementioned limitations of current discrete models.

\begin{figure*}[htbp!]
\centering\includegraphics*[width = \linewidth]{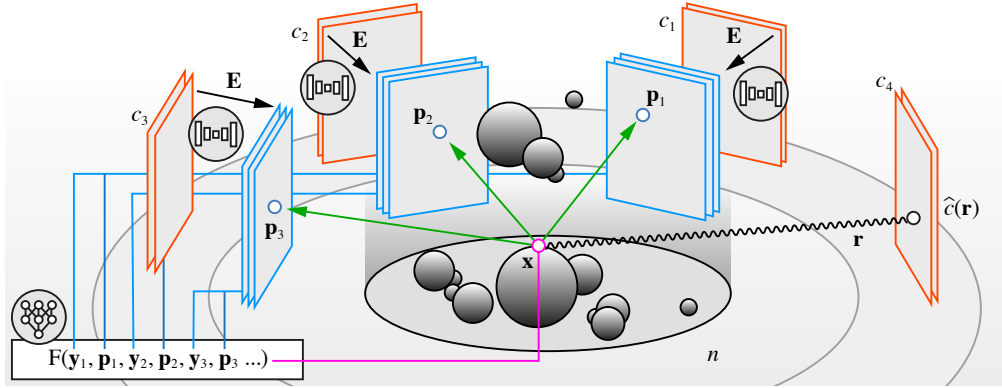}
\caption{Overall scheme of \name. 
Four training views ($\contrast_1$,...,$\contrast_4$) are shown in the figure, where the first three are selected as constraints, i.e., views to condition the 3D model.
The views used as constraints are encoded into latent space via a encoder (\encoder).
For each 3D spatial coordinate \worldSpaceCoord, the predicted complex index of refraction (\ior) can be related to the local coordinates ($\viewSpaceCoord_1$, $\viewSpaceCoord_2$, $\viewSpaceCoord_3$) of the encoded views and the corresponding latent vectors ($\prior_1$, $\prior_2$, $\prior_3$) via a fully connected neural network (\mlp).
The function between the 3D complex index of refraction (\ior) and the 3D spatial coordinate (\worldSpaceCoord) is learned by optimizing the difference between the pixel value of the input image ($\contrast_4(\ray)$) and the predicted line-integral of \ior ($\widehat{c}(\ray)$) along each ray (\ray).}
\label{fig:main}
\end{figure*}

\name is an implicit neural representation that retrieves the 3D complex index of refraction (\ior) as a function of the 3D spatial coordinate \worldSpaceCoord from a sparse number of views. 
\name has two main components, and its implementation is depicted in Fig.~\ref{fig:main}. 

The first part is an encoder (\encoder) based on a 2D \ac{CNN}
that captures the information across the measured 2D views for different stereoscopic or X-ray multi-projection experiments. 
Thus, $\encoder$ enables learning 3D objects across experiments in order to enhance the 3D information retrieved from sparse-view approaches.
In each training iteration, we randomly encode a subset of the projections (\priorView) from all measured ones (\allKnownView).
We call the subset of encoded views constraints, as they are used to constrain our 3D volume.
\encoder takes as input the attenuation and phase-contrast images of each of the \viewIndex-th views in that subset ($\contrast_{\priorView}$) and encodes them into a high-dimensional space also known as latent space ($\encoder(\contrast_\viewIndex)$).
The latent space provided by the $\encoder(\contrast_\viewIndex)$ can be understood as a pixel image where every pixel contains a vector, also known as the latent vector, instead of a scalar as in a conventional single-channel image.
To obtain the encoded information for any given 3D point \worldSpaceCoord, we use the affine coordinate transformation \worldToViewSpace to transfer from the global 3D coordinate system to the local coordinate system of each view, where the point coordinate of \worldSpaceCoord in the view \viewIndex local system is denoted by $\viewSpaceCoord_\viewIndex=\worldToViewSpace_\viewIndex(\worldSpaceCoord)$.
In the local coordinate system for each view or detector position, the 3D volume is parametrized in terms of i) the distance between any point and the detector plane along the propagation direction of the X-rays that generated that view and ii) the intersection coordinates of that ray with the camera plane.
This coordinate system simplifies the projection of information between the encoded images and the object information in 3D. 
Our encoder \encoder is based on ResNet34~\cite{he2016deep}, which has three convolutional operations to extract key features and three pooling layers to reduce the dimension of the convolution results to keep the translation invariance of the image.
The second component uses the information provided by each $\encoder(\contrast_\viewIndex)$ and the geometrical information of the experimental arrangement.

The second component (\mlp) is the main neural network.
It reconstructs an object ($\ior(\mathbf x)$) from the given constraints and prior information using a fully-connected neural network.
Specifically, we use a fully-connected neural network with five layers, width 128, and residual connections \cite{he2016deep}, similar to the ones in Ref.~\cite{yu2021pixelnerf}. 
This implementation is composed of three parts.
The first part of \mlp has \priorView parallel weight-sharing ResNet 3$\times$ blocks.
Each of these parallel blocks takes, for a spatial point \worldSpaceCoord, the converted local coordinates $\viewSpaceCoord_\viewIndex$ from all constraints, and the corresponding latent vectors $\prior_\viewIndex[\viewSpaceCoord_\viewIndex]$ that contain the projected information of  \worldSpaceCoord into that view.
Note that for each 3D spatial point, we only need the pixel-aligned latent vector from each constraint $\prior_\viewIndex[\viewSpaceCoord_\viewIndex]$, not all the encoded information per view $\encoder(\contrast_\viewIndex)$.
The second part is just an average operator, which takes the mean of the processed information from the different views.
The third part is made out of two ResNet blocks, introducing new learnable parameters to the averaged information. 

Implementations, like the one used by \name, are known to be biased towards learning low-resolution or low-frequency functions~\cite{rahaman2019spectral}. 
It has also been demonstrated that mapping the inputs to a higher dimensional space allows the networks to learn better high-frequency functions~\cite{tancik2020fourier}.
For this reason, the coordinates are mapped from $\mathbb{R}$ to a high-dimension space $\mathbb{R}^{2L}$ using positional encoding~\cite{vaswani2017attention,mildenhall2020nerf}. The positional encoding can be expressed in terms of a Fourier basis as:
\begin{equation}
    \posionalEncoder(\worldSpaceCoord)=\left(\sin \left(2^{0} \pi \worldSpaceCoord\right), \cos \left(2^{0} \pi \worldSpaceCoord\right), \cdots, \sin \left(2^{L-1} \pi \worldSpaceCoord\right), \cos \left(2^{L-1} \pi \worldSpaceCoord\right)\right)~,
    \label{Eq: positional_encoding}
\end{equation}
where $L$ is a positive integer. In this work, we set $L$ = 10.
As a result, we express for each \worldSpaceCoord a generated $\ior$ or output of \name as:

\begin{equation}
    \ior(\worldSpaceCoord) =
    \mlp(
    \posionalEncoder(\viewSpaceCoord_1),
    \prior_1[\viewSpaceCoord_1],
    \ldots,
    \posionalEncoder(\viewSpaceCoord_\viewIndex),
    \prior_\viewIndex[\viewSpaceCoord_m],
    \ldots,
    \posionalEncoder(\viewSpaceCoord_\priorView),
    \prior_\priorView[\viewSpaceCoord_\priorView]    
    ).
\end{equation}

With the 3D information of the refractive index $\ior(\mathbf x)$, we can calculate the 2D projection images at any viewpoint.
This transfer between 3D and 2D is done via ray tracing and the projection approximation as described in subsection~\ref{subsec:interaction}. 
In this work, we deal with a fixed distance between the cameras and the sample and a parallel beam geometry, i.e., all the X-rays that generate a view have the same direction.
However, this approach can be adapted quickly to other experimental setups, such as the cone-beam or fan-beam configurations, by modifying the geometric expressions that described the experimental configuration in our \href{https://github.com/pvilla/ONIX.git}{code}.
We parametrize the rays as $\ray(\thickness) = \rayOrigin + \thickness\rayDirection$,  where $\thickness$ denotes the coordinate along the ray, $\rayOrigin$ and $\rayDirection$ stand for ray origin and ray direction, respectively. 
Under the parallel-beam geometry, the origins correspond to the position of sensor pixels, and the ray directions are the same for each individual view.
%
The 2D calculated rays and projections from the constraints are used to optimize the parameters of \name using the measured views ($\contrast_{\allKnownView}$).
This is done by optimizing the following loss function:
\begin{equation}
\mathcal L = \sum_{\ray \in \allRays} \left \|\contrast_{\allKnownView}(\ray) - \genintensity(\ray) \right \|^2_2,
\label{Eq:loss}
\end{equation}
where \allRays is the set of all rays and $\genintensity$ denotes the estimated phase or attenuation contrast, respectively.
Due to computational limitations, we do not calculate the loss function on all of the pixel points or all possible rays of an image in the actual training.
Instead, we select a set of ray sampling points or pixels from all input views and calculate the loss function on these sampling points.
The rays are selected randomly over a probability density function based on the magnitude of the spatial gradient of the input images. 
The calculated pixel values for phase and attenuation are obtained using a discrete version of the projection approximation:
\begin{equation}
    \genintensity = \log(\psi_z / \psi_{z_0}) = \sum_{j=1}^{N}\distBetweenAdjacentSample_j(- \imag k\realIOR_j - k\imagIOR_j),
    \label{Eq: ray_rendering}
\end{equation}
where $\distBetweenAdjacentSample_j = \thickness_{j+1}-\thickness_j$ is the distance between two adjacent sampling points.
The sampling points used along the ray are selected randomly.
By calculating the squared $L_2$ norm between the line integral of the predicted \ior and the pixel value from the inputs, \name can be trained to learn 3D self-consistency between different projections.

After the training, \name can generate 3D reconstructions using only the subset of constraints ($\priorView$). 
As shown in Fig.~\ref{fig:main}, if we had four known views for each object, we could choose three of them (denoted by indices 1-3 in Fig.~\ref{fig:main}) as constraints ($\priorView$) to encode their information.
Please note that one can use an arbitrary number (smaller than the number of total input views) of views, as few as one, as the subset of constraints, depending on the actual situation.
Having more constraints in training provides better results and increases computational demand.
In this work, given that we had eight known views in total, we constrained our 3D reconstructions with four and six encoded views for the simulated and experimental datasets, respectively.

\name was implemented in PyTorch 1.6.0 and Python 3.8.8.
The training was performed on an NVIDIA A100 GPU with 40~GB of RAM.
For training of the simulation dataset, we chose 1024 ray sampling points from all input views of an object to calculate the loss along each ray.
For each ray, 256 depth points were sampled.
The number of sampling points can be flexibly adjusted based on hardware conditions.
The Adam optimization algorithm \cite{kingma2014adam} with a mini-batch size of two was used for the learning. 
We divided the training dataset into mini-batches because the dataset was too large to be loaded into the memory all at a time.
We used a batch size of two, meaning that each mini-batch contained all known views of two objects or independent experiments, and the networks needed 500 training iterations to load all 1000 objects and finish an epoch, i.e., one complete training cycle. 
The step size of the optimizer, or the learning rate, was set to be 0.005 for the networks. 
We stopped the training after 500 epochs. 
This took \textasciitilde 24 hours.
For the training of the experimental dataset, we selected 3096 ray sampling points from all input views, and 64 depth samples were chosen along each ray. 
As we only had attenuation contrast, the phase-contrast channel was removed from the output of the \mlp, i.e., we had a single output channel for the attenuation reconstruction.
We used the same initial learning rate as for the simulation dataset but reduced the learning rate by 0.1 after 1000 epochs.
The training was stopped after 1500 epochs, which took \textasciitilde 12 hours.

\subsection{3D supervised learning\label{subsec:supervised}}
This subsection introduces a 3D-supervised baseline  \ac{DL} approach, which is used to assess the capabilities of \name, our non-3D unsupervised approach, in comparison to a \ac{DL} approach that has access to 3D information. %
The 3D-supervised approach retrieved volumetric information by training and reconstructing stacks of individual 2D slices~\cite{Liu2020TomoGAN}.
The paired dataset used for this supervised approach consisted of: i) input: 2D slices reconstructed from only eight views using \ac{SART}, and ii) ground truth: 2D slices obtained either from the simulated volume or reconstructed from 96 views for the experimental data. 
Further details of the datasets and the data preparation can be found in subsections~\ref{subsec:SimulatedData} and \ref{subsec:ExperimentalData}.
The architecture of the \ac{DL} was based on \ac{CNN}s given their unique capabilities for image reconstruction and pattern recognition.
Specifically, we used a U-NET~\cite{ronneberger2015u}, a \ac{CNN} architecture that includes an image contracting path consisting of several downblocks, and an image expanding path consisting of the same amount of upblocks. 
In our task, the size of the input and output datasets was the same ($256 \times 256$). 
The optimization or loss function for this approach was an \ac{MSE} or $L_2$ norm between the ground-truth and the output of the U-NET.
In order to minimize the overfitting, a $L_2$ regularization is implemented during the minimization process to penalize large values of the weights or fitted parameters of the U-NET architecture.
The training of the U-NET was also performed on PyTorch 1.6.0 and Python 3.8.8 using an NVIDIA A100 GPU with 40 GB of RAM.     

We should highlight that there were some slight differences between the training of the simulated and experimental data.
First, the number of channels was different. 
The U-NET had two input channels for the simulated data as the attenuation and phase-information were available. 
For the experimental data, one channel was used, as only the attenuation channel was available. 
Second, the architecture or structure of the U-NET was different.
For the simulated data, the two-channel U-NET started with 16 filters and contained a total of five downblocks and five upblocks. 
For the experimental data, the one-channel U-NET started with 32 filters and contained a total of six downblocks and six upblocks.
Finally, the hyperparameters used in training were different. 
For the simulated data, the $L_2$-regularization was set as 0.001. 
For experimental data, the $L_2$-regularization was set as 0.005.    

\subsection{Data analysis}
This subsection describes the approaches used to evaluate the performance of the different 3D reconstruction algorithms from a sparse number of projections: i) supervised 3D learning, ii) \ac{SART}, and iii) our approach \name. 
We used two datasets to validate and analyze the capabilities of the three different approaches: i) simulated tomographic data using ellipsoid objects, described in subsection~\ref{subsec:SimulatedData}, and ii) experimental data using metallic foams as described in subsection~\ref{subsec:ExperimentalData}.
In this work, we used only eight projections to train and retrieve our 3D reconstructed volumes for all the reconstruction approaches presented.
The reconstructed 3D volumes were compared to either the simulated 3D volume or the reconstructed volume from a large number of projections for the simulated and experimental datasets, respectively.
This comparison was done using two different image quality metrics: $L_2$ norm and \ac{DSSIM}, as shown in Table \ref{tbl:thixo_results}.
These two metrics calculate the image correlation in the object space and better quality corresponds to smaller values of these two metrics.
We applied a ring mask to remove the borders and keep only the relevant information for comparison purposes to all the reconstructed volumes and slices.
This is a common practice to evaluate the performance of tomographic and 3D reconstruction algorithms.
The same mask was used to produce the ground truth and the DL methods results.
Furthermore, for the experimental dataset, we set a cutoff value for the masked outputs to select the region of interest for validation and improved visualization in the reference volumes.
No threshold was applied to the \ac{SART} reconstruction.
For the 3D rendering, we applied a common threshold to the 3D reconstructions to visualize the metallic foams. 
Different thresholds were applied to the \ac{SART} reconstructions to show the feature of interest for the both simulation and experimental dataset. 

\section{Results and discussion}
This section presents the results of \name when applied to simulated (subsection~\ref{subsec:SimulationResults}) and experimental data (subsection~\ref{subsec:ExperimentalResults}).
For comparison purposes, we studied the reconstructions using state-of-the-art iterative reconstructions based on \ac{SART} and a supervised 3D approach described in subsection~\ref{subsec:supervised}.
For simplicity, we refer in this section to the three methods as 3D supervised, \ac{SART}, and \name.
Finally, we discussed all the obtained results in the last subsection (subsection~\ref{subsec:ResultsDiscussion}).

\subsection{Simulation results}
\label{subsec:SimulationResults}

\begin{figure*}[htbp!]
\centering\includegraphics*[width = \linewidth]{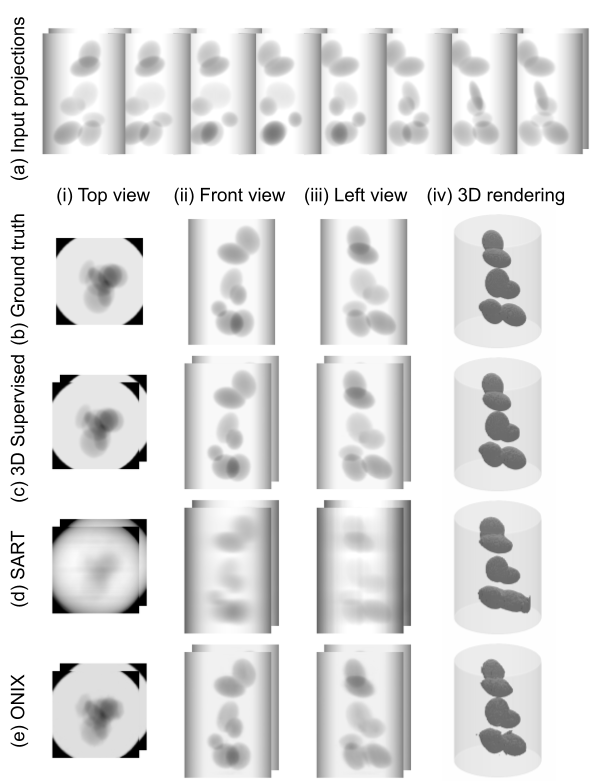}
\caption{
Tomographic reconstruction results for the simulation dataset. (a) The eight input projections that were used for the reconstructions. (b) The simulated ground truth. (c-e) Reconstruction results from (c) 3D supervised learning, (d) SART, and (e) \name. The input projections and the output results have both absorption- and phase-contrast components. We show only the absorption contrast results overlaid with the phase. For the ground truth and the three reconstruction methods, we depict the projections along an orthogonal system from the (i) top view, (ii) front view, and (iii) and left view, as well as the (iv) 3D rendered reconstructions.
}

\label{fig:ellipsoid_result}
\end{figure*}
In this subsection, we applied \name on the simulated tomographic data using ellipsoid objects and compared the performance of \name with \ac{SART} and 3D supervised learning. 

Fig.~\ref{fig:ellipsoid_result} shows the inputs and outputs of the compared tomographic reconstruction approaches of a validation object, together with its ground truth.
The validation object was not in the training dataset of the supervised learning, while its projections were in the training dataset of \name.
As discussed in subsection~\ref{subsec:SimulatedData}, we used eight projection angles for the training of the supervised, and \name approaches.
On the one hand, the supervised approach also accessed the simulated 3D volume in the training process as shown in Fig~\ref{fig:Concept}.
On the other hand, \name and \ac{SART} are fully unsupervised, i.e., no 3D information was accessed in any stage of learning or reconstruction, see Fig.~\ref{fig:Concept}. 
The \name results shown here were reconstructed using four views as contraints of the eight recorded projections.
For the specific case of the \ac{SART} reconstructions, we used the same eight projections used for the training and reconstructions.
The projected absorption- and phase-contrast images used to obtain the depicted results are shown in Fig.~\ref{fig:ellipsoid_result} (a).
To visually evaluate the reconstructions from the three different algorithms, we plotted for each reconstruction algorithm the projections of the 3D volume through three orthogonal axes denoted by (i) top, (ii) front, and (iii) left.
We also rendered the 3D reconstructed volumes for all the algorithms and ground truth (iv).
The results of these four evaluation images (i-iv) for the ground truth, 3D supervised, \ac{SART}, and \name are shown in Fig~\ref{fig:ellipsoid_result}(b-e), respectively.
The absorption and phase reconstructions were very similar as the ratio of the real and imaginary parts of the index of refraction for a single material simulation were fixed, see subsection~\ref{subsec:SimulatedData}.
For simplicity, we showed only the absorption contrast in Fig.~\ref{fig:ellipsoid_result}.
These results can be compared with the ground truth: simulated 3D objects. 
To quantitatively evaluate the 3D reconstructions from the three methods, we computed the $L_2$ norm and \ac{DSSIM} of the results in Table~\ref{tbl:thixo_results}. 
We validated the absorption- and phase-contrast volumes separately and then took the average of the two. 
A reader is referred to the supplementary material to compare the results over slices.

\begin{table}[htbp]
\centering
\caption{Comparison of reconstructed results using different methods}
\label{tbl:thixo_results}
\begin{tabular}{ccccc}
\hline
 & \multicolumn{2}{c}{Simulation dataset} & \multicolumn{2}{c}{Experimental dataset} \\
Reconstructed results                      & $L_2$  \footnotesize $\times 10^{-2}$ & DSSIM  \footnotesize $\times 10^{-2}$   & $L_2$\footnotesize $\times 10^{-2}$ & DSSIM \footnotesize $\times 10^{-2}$ \\ \hline
3D supervised                             &  1.1          &  0.86                & 5.5   & 3.3        \\
SART                                      &  21          &     7.5               & 21  &   16     \\ 
\name                                     &  5.1         &     1.6               & 5.4    &    3.3      \\ \hline
\end{tabular}
\end{table}

\subsection{Experimental result}
\label{subsec:ExperimentalResults}
To demonstrate the capability of \name, we also applied it to the experimental metallic foam dataset collected from TOMCAT, an X-ray imaging instrument at the Swiss Light Source, as described in \ref{subsec:ExperimentalData}.
The collected images in this experiment only contained absorption-contrast as the coherence of the experiment was not sufficient to retrieve phase-contrast information.

\begin{figure*}[htbp!]
\centering\includegraphics*[width = \linewidth]{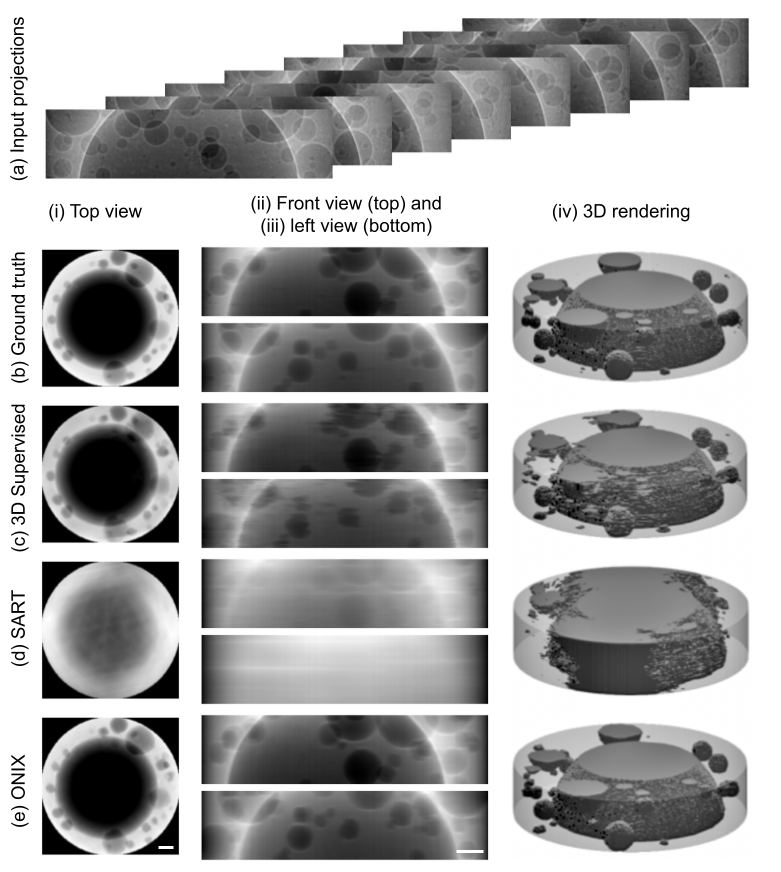}
\vspace{-.2cm}
\caption{Tomographic reconstruction results from different methods. (a) The eight input projections used for training and reconstructions. (b) The ground truth obtained from SART using 96 projections. (c-e) Reconstruction results from (c)3D supervised learning, (d) SART, and (e)\name. The (i) top view, (ii) front view, (iii) left view, and (iv) 3D rendering images are plotted for the ground truth volume and the 3D reconstructions. The size of the top views was adjusted for better arrangement. The scale bar shown in (e) corresponds to 200 $\mu$m.}

\label{fig:thixo_result}
\end{figure*}

Fig.~\ref{fig:thixo_result} shows (a) measured absorption-contrast images at eight equally-spaced projection angles, between 0$\degree$ and 131$\degree$ of the metallic foam, (b) \ac{SART}-retrieved ground truth from 96 projections, and the corresponding reconstruction results from (c) 3D supervised learning, (d) \ac{SART}, and (e) \name using six encoded views. 
Similarly, we plotted in different columns the (i) top, (ii) front, (iii) left projections through orthogonal axes, and (iv) side-view rendering of the 3D reconstructions. 
The results were rendered from the projection angle of 169$\degree$, which was not used for our reconstructions.
The reader is referred to the supplementary material that visualizes the results of the reconstructions over orthogonal slices and provides a movie showing the comparison between the ground truth and the three reconstruction results. 
It can be seen from Fig.~\ref{fig:thixo_result} that under our experimental conditions, \ac{SART} with 96 projections can provide clean and sharp 3D reconstructions, which is reliable enough for the supervised learning to learn from and for the other methods to be compared. 
The quantitative assessment for the experimental dataset is also shown in Table~\ref{tbl:thixo_results}.
In the supplementary material, we also compared the \name reconstruction results using four and six views to reconstruct our 3D volume and the results from all the different combinations of them.
The results shown in Fig. \ref{fig:thixo_result}(e) were the best reconstructions we found using six views among our investigation.

\subsection{Discussion}
\label{subsec:ResultsDiscussion}

This subsection studies the results presented in \ref{subsec:SimulationResults} and \ref{subsec:ExperimentalResults} to analyze the capabilities of \name by comparing it with state-of-the-art SART approach and a 3D supervised learning approach. 
It can be seen from Table \ref{tbl:thixo_results} and the figures that, generally speaking, 3D supervised learning provides the most competitive results among the three experimented approaches.
However, this approach requires accessing the 3D ground truth during training, limiting its applicability to techniques and scenarios where it is possible to acquire 3D information.
Thus, this approach can be conceived as a simple denoising approach rather than an enabling technique for sparse acquisition approaches.
On the other hand, \name taking no 3D supervision i) clearly outperforms state-of-the-art iterative reconstruction like \ac{SART} using the same information, and ii) performs almost at the same level as the 3D supervised approach. 
If one compares the simulated and experimental results, it is noticeable that the 3D supervised learning results for the experimental dataset have inferior quality to the results of the simulations.
This difference may arise from two factors.
First, this could partially be because the ground truth is downsized due to the memory limitation of the computing resources.
Thus, the benefit of \name having infinite resolution becomes apparent, as the downsizing of the input projections has less impact than it does for 3D supervised learning.
Second, the noise contribution of the simulated and experimental data are unrelated.
The performance of \name depends on the number of available views. 
Although \name implementation can use a random number of input views and as low as one, we have focused on eight views for this article based on current designs for X-ray multi-projection imaging techniques~\cite{villanueva2018hard,Voegeli2020XMPI}.
We have observed that even using the same number of projections, \name provides better results for the experimental data when six views are used instead of four, as shown in the supplementary material.
Thus, the performance of \name can be further improved by encoding with more views given the same amount of projections.
We have also found that the performance of \name is independent of the permutation of views used as constraints but dependent on the combination of them.
In general, the reconstruction obtained from closely spaced projection angles is not as good as the ones obtained from widely spaced angles.
The best reconstructions were obtained when choosing widely spaced projections as constraints, which is the case for results shown in \ref{subsec:SimulationResults} and \ref{subsec:ExperimentalResults}.
Although we have only applied \name on objects with known 3D information, we have demonstrated its capabilities and potential when using a subset of sparse projections that emulates realistic conditions for X-ray multi-projection imaging experiments. 
The unique capabilities of \name to enhance 3D reconstructions to levels previously impossible from only a sparse set of views are the result of including the physics of image formation with X-rays and transferring prior knowledge between different objects and experiments~\cite{yu2021pixelnerf,henzler2021unsupervisedvideos}.
We envision that \name will become an enabling tool for 3D sparse X-ray imaging techniques like time-resolved X-ray multi-projection imaging experiments and medical stereoscopy.

\section{Conclusion}
To conclude, we have presented \name, a 3D unsupervised DL approach for X-ray imaging that can retrieve high-quality 3D reconstructions from an arbitrary set of sparse projections.
\name extends the capabilities of current 3D volume reconstruction approaches from sparse techniques by i) offering an implicit representation with a theoretically infinite resolution that overcomes the memory demands of approaches like conventional voxel representations, ii) including the physical description of the interaction of X-rays with matter, enables an accurate transfer of the information between the 3D model and the recorded 2D projections, and iii) including a convolutional neural network that transfers prior knowledge between experiments and samples to enhance the performance of this approach.
\name can retrieve, in general, the complex index of refraction, making our approach not only compatible with traditional absorption-contrast radiographs but also phase-contrast or coherent imaging techniques. 
To demonstrate the capabilities of \name, we have compared it to state-of-the-art iterative reconstructions (\ac{SART}) and 3D supervised machine learning approaches using only eight views under realistic conditions for X-ray multi-projection imaging experiments. 
We have observed with simulated and experimental data that \name outperforms current iterative approaches and performs at the level of 3D supervised approaches.
However, 3D supervised approaches are only presented for comparison purposes as these methods cannot be applied to experiments with sparse 2D views as the 3D information required for the training is not available.
Therefore, we envision \name as a tool that will enable 3D reconstructions for X-ray imaging at arbitrary resolutions from a sparse set of views with a quality that was not previously possible.
Specifically, we envision its application to expand i) the capabilities of time-resolved experiments not possible today that exploit the unique properties and high-flux of XFELs and diffraction-limited storage rings, and ii) the information provided by stereoscopic techniques for different fields such as medical applications.

The \name code is available at \href{https://github.com/pvilla/ONIX.git}{GitHub}. 
\begin{backmatter}

\bmsection{Acknowledgments}
We thank F. Garcia-Moreno, Mike Noack, and Paul H. Kamm from the Helmholtz-Zentrum Berlin for providing the metallic foam data used for the experimental validation and insightful discussions about this data.
We wish to acknowledge the insightful comments of Elizabeth Blackburn and Gary Harlow.
We are grateful to Z. Matej for his support and access to the GPU-computing cluster at MAX IV. 
We are also grateful to the European XFEL for its support, especially the computing and data analysis groups, for granting access to the GPU-computing cluster.
We also gratefully acknowledge the support of NVIDIA Corporation with the donation of a Quadro P4000 GPU used for this research.

\bmsection{Disclosures}
The authors declare no conflicts of interest.

\bmsection{Data availability} Data underlying the results presented in this paper are not publicly available at this time but may be obtained from the authors upon reasonable request.

\bmsection{Supplemental document}
See Supplement 1 for supporting content.

\end{backmatter}

\bibliography{Article}

\begin{thebibliography}{10}
\newcommand{\enquote}[1]{``#1''}

\bibitem{Cormack1964CT}
A.~M. Cormack, \enquote{{Representation of a function by its line integrals,
  with some radiological applications. II},} {\protect\JournalTitle{Journal of
  Applied Physics}} \textbf{35}, 2908--2913 (1964).

\bibitem{Hounsfield1973CT}
G.~N. Hounsfield, \enquote{{Computerized transverse axial scanning
  (tomography): I. Description of system},} {\protect\JournalTitle{British
  Journal of Radiology}} \textbf{46}, 1016--1022 (1973).

\bibitem{Davidovits1969Confocal}
P.~Davidovits and M.~D. Egger, \enquote{{Scanning laser microscope},}
  {\protect\JournalTitle{Nature}} \textbf{223}, 831 (1969).

\bibitem{Ale2012ConfocalFluorescence}
A.~Ale, V.~Ermolayev, E.~Herzog, C.~Cohrs, M.~H. de~Angelis, and
  V.~Ntziachristos, \enquote{{FMT-XCT: in vivo animal studies with hybrid
  fluorescence molecular tomography–X-ray computed tomography},}
  {\protect\JournalTitle{Nature Methods}} \textbf{9}, 615--620 (2012).

\bibitem{Salvo2003CTinMS}
L.~Salvo, P.~Cloetens, E.~Maire, S.~Zabler, J.~J. Blandin, J.~Y.
  Buffi{\`{e}}re, W.~Ludwig, E.~Boller, D.~Bellet, and C.~Josserond,
  \enquote{{X-ray micro-tomography an attractive characterisation technique in
  materials science},} in \emph{Nuclear Instruments and Methods in Physics
  Research, Section B: Beam Interactions with Materials and Atoms,}  vol. 200
  (2003), pp. 273--286.

\bibitem{Galban2012CTMedicine}
C.~J. Galb{\'{a}}n, M.~K. Han, J.~L. Boes, K.~A. Chughtai, C.~R. Meyer, T.~D.
  Johnson, S.~Galb{\'{a}}n, A.~Rehemtulla, E.~A. Kazerooni, F.~J. Martinez, and
  B.~D. Ross, \enquote{{Computed tomography–based biomarker provides unique
  signature for diagnosis of COPD phenotypes and disease progression},}
  {\protect\JournalTitle{Nature Medicine}} \textbf{18}, 1711--1715 (2012).

\bibitem{Ale2012CTanimals}
A.~Ale, V.~Ermolayev, E.~Herzog, C.~Cohrs, M.~H. de~Angelis, and
  V.~Ntziachristos, \enquote{{FMT-XCT: in vivo animal studies with hybrid
  fluorescence molecular tomography–X-ray computed tomography},}
  {\protect\JournalTitle{Nature Methods}} \textbf{9}, 615--620 (2012).

\bibitem{Withers2021CTreview}
P.~J. Withers, C.~Bouman, S.~Carmignato, V.~Cnudde, D.~Grimaldi, C.~K. Hagen,
  E.~Maire, M.~Manley, A.~{Du Plessis}, and S.~R. Stock, \enquote{{X-ray
  computed tomography},} {\protect\JournalTitle{Nature Reviews Methods
  Primers}} \textbf{1}, 18 (2021).

\bibitem{Garcia-Moreno2019FastCT}
F.~Garc{\'{i}}a-Moreno, P.~H. Kamm, T.~R. Neu, F.~B{\"{u}}lk, R.~Mokso, C.~M.
  Schlep{\"{u}}tz, M.~Stampanoni, and J.~Banhart, \enquote{{Using X-ray
  tomoscopy to explore the dynamics of foaming metal},}
  {\protect\JournalTitle{Nature Communications}} \textbf{10}, 3762 (2019).

\bibitem{Ziesche2020FastCTBatteries}
R.~F. Ziesche, T.~Arlt, D.~P. Finegan, T.~M.~M. Heenan, A.~Tengattini, D.~Baum,
  N.~Kardjilov, H.~Mark{\"{o}}tter, I.~Manke, W.~Kockelmann, D.~J.~L. Brett,
  and P.~R. Shearing, \enquote{{4D imaging of lithium-batteries using
  correlative neutron and X-ray tomography with a virtual unrolling
  technique},} {\protect\JournalTitle{Nature Communications}} \textbf{11}, 777
  (2020).

\bibitem{Evans2003StereoDynamic}
J.~Evans and H.~Hon, \enquote{Dynamic stereoscopic x-ray imaging,}
  {\protect\JournalTitle{NDT \& E International}} \textbf{35}, 337--345 (2002).

\bibitem{Verellen2003StereoMedical}
D.~Verellen, G.~Soete, N.~Linthout, S.~{Van Acker}, P.~{De Roover},
  V.~Vinh-Hung, J.~{Van de Steene}, and G.~Storme, \enquote{Quality assurance
  of a system for improved target localization and patient set-up that combines
  real-time infrared tracking and stereoscopic x-ray imaging,}
  {\protect\JournalTitle{Radiotherapy and Oncology}} \textbf{67}, 129--141
  (2003).

\bibitem{Wang2006LowdoseCT}
J.~Wang, T.~Li, H.~Lu, and Z.~Liang, \enquote{Penalized weighted least-squares
  approach to sinogram noise reduction and image reconstruction for low-dose
  x-ray computed tomography,} {\protect\JournalTitle{IEEE Transactions on
  Medical Imaging}} \textbf{25}, 1272--1283 (2006).

\bibitem{Zanette2012CTLow-dose}
I.~Zanette, M.~Bech, A.~Rack, G.~Le~Duc, P.~Tafforeau, C.~David, J.~Mohr,
  F.~Pfeiffer, and T.~Weitkamp, \enquote{Trimodal low-dose x-ray tomography,}
  {\protect\JournalTitle{Proceedings of the National Academy of Sciences}}
  \textbf{109}, 10199--10204 (2012).

\bibitem{thomson1896stereoscopic}
E.~Thomson, \enquote{Stereoscopic roentgen pictures,}
  {\protect\JournalTitle{Electr Eng}} \textbf{2} (1896).

\bibitem{Howells1994XMPI}
M.~R. Howells and C.~J. Jacobsen, \enquote{{Possibility for one-shot tomography
  using a high-gain free-electron laser},} {\protect\JournalTitle{Workshop on
  scientific applications of coherent X-rays}} \textbf{SLAC-R-437}, 159--162
  (1994).

\bibitem{villanueva2018hard}
P.~Villanueva-Perez, B.~Pedrini, R.~Mokso, P.~Vagovic, V.~A. Guzenko, S.~J.
  Leake, P.~R. Willmott, P.~Oberta, C.~David, H.~N. Chapman \emph{et~al.},
  \enquote{Hard x-ray multi-projection imaging for single-shot approaches,}
  {\protect\JournalTitle{Optica}} \textbf{5}, 1521--1524 (2018).

\bibitem{Moseley1962StereoXrayFirstDevice}
R.~D. Moseley and J.~R. Williams, \enquote{Apparatus for stereoscopic chest
  radiography,} {\protect\JournalTitle{Radiology}} \textbf{78}, 103--108
  (1962). PMID: 14476439.

\bibitem{Eriksson2014DLSR}
M.~Eriksson, J.~F. van~der Veen, and C.~Quitmann, \enquote{{Diffraction-limited
  storage rings {--} a window to the science of tomorrow},}
  {\protect\JournalTitle{Journal of Synchrotron Radiation}} \textbf{21},
  837--842 (2014).

\bibitem{McNeil2010XFELreview}
B.~W.~J. McNeil and N.~R. Thompson, \enquote{{X-ray free-electron lasers},}
  {\protect\JournalTitle{Nature Photonics}} \textbf{4}, 814--821 (2010).

\bibitem{Emma2010XFEL_operation}
P.~Emma, R.~Akre, J.~Arthur, R.~Bionta, C.~Bostedt, J.~Bozek, A.~Brachmann,
  P.~Bucksbaum, R.~Coffee, F.~J. Decker, Y.~Ding, D.~Dowell, S.~Edstrom,
  A.~Fisher, J.~Frisch, S.~Gilevich, J.~Hastings, G.~Hays, P.~Hering, Z.~Huang,
  R.~Iverson, H.~Loos, M.~Messerschmidt, A.~Miahnahri, S.~Moeller, H.~D. Nuhn,
  G.~Pile, D.~Ratner, J.~Rzepiela, D.~Schultz, T.~Smith, P.~Stefan,
  H.~Tompkins, J.~Turner, J.~Welch, W.~White, J.~Wu, G.~Yocky, and J.~Galayda,
  \enquote{{First lasing and operation of an {\aa}ngstrom-wavelength
  free-electron laser},} {\protect\JournalTitle{Nature Photonics}} \textbf{4},
  641--647 (2010).

\bibitem{SACLA2012Huang}
Z.~Huang and I.~Lindau, \enquote{{SACLA hard-X-ray compact FEL},}
  {\protect\JournalTitle{Nature Photonics}} \textbf{6}, 505--506 (2012).

\bibitem{Kang2017PALXFEL}
H.-S. Kang, C.-K. Min, H.~Heo, C.~Kim, H.~Yang, G.~Kim, I.~Nam, S.~Y. Baek,
  H.-J. Choi, G.~Mun, B.~R. Park, Y.~J. Suh, D.~C. Shin, J.~Hu, J.~Hong,
  S.~Jung, S.-H. Kim, K.~Kim, D.~Na, S.~S. Park, Y.~J. Park, J.-H. Han, Y.~G.
  Jung, S.~H. Jeong, H.~G. Lee, S.~Lee, S.~Lee, W.-W. Lee, B.~Oh, H.~S. Suh,
  Y.~W. Parc, S.-J. Park, M.~H. Kim, N.-S. Jung, Y.-C. Kim, M.-S. Lee, B.-H.
  Lee, C.-W. Sung, I.-S. Mok, J.-M. Yang, C.-S. Lee, H.~Shin, J.~H. Kim,
  Y.~Kim, J.~H. Lee, S.-Y. Park, J.~Kim, J.~Park, I.~Eom, S.~Rah, S.~Kim, K.~H.
  Nam, J.~Park, J.~Park, S.~Kim, S.~Kwon, S.~H. Park, K.~S. Kim, H.~Hyun, S.~N.
  Kim, S.~Kim, S.-m. Hwang, M.~J. Kim, C.-y. Lim, C.-J. Yu, B.-S. Kim, T.-H.
  Kang, K.-W. Kim, S.-H. Kim, H.-S. Lee, H.-S. Lee, K.-H. Park, T.-Y. Koo,
  D.-E. Kim, and I.~S. Ko, \enquote{{Hard X-ray free-electron laser with
  femtosecond-scale timing jitter},} {\protect\JournalTitle{Nature Photonics}}
  \textbf{11}, 708--713 (2017).

\bibitem{Prat2020SwissFEL}
E.~Prat, R.~Abela, M.~Aiba, A.~Alarcon, J.~Alex, Y.~Arbelo, C.~Arrell,
  V.~Arsov, C.~Bacellar, C.~Beard, P.~Beaud, S.~Bettoni, R.~Biffiger, M.~Bopp,
  H.-H. Braun, M.~Calvi, A.~Cassar, T.~Celcer, M.~Chergui, P.~Chevtsov,
  C.~Cirelli, A.~Citterio, P.~Craievich, M.~C. Divall, A.~Dax, M.~Dehler,
  Y.~Deng, A.~Dietrich, P.~Dijkstal, R.~Dinapoli, S.~Dordevic, S.~Ebner,
  D.~Engeler, C.~Erny, V.~Esposito, E.~Ferrari, U.~Flechsig, R.~Follath,
  F.~Frei, R.~Ganter, T.~Garvey, Z.~Geng, A.~Gobbo, C.~Gough, A.~Hauff, C.~P.
  Hauri, N.~Hiller, S.~Hunziker, M.~Huppert, G.~Ingold, R.~Ischebeck,
  M.~Janousch, P.~J.~M. Johnson, S.~L. Johnson, P.~Jurani{\'{c}}, M.~Jurcevic,
  M.~Kaiser, R.~Kalt, B.~Keil, D.~Kiselev, C.~Kittel, G.~Knopp, W.~Koprek,
  M.~Laznovsky, H.~T. Lemke, D.~L. Sancho, F.~L{\"{o}}hl, A.~Malyzhenkov, G.~F.
  Mancini, R.~Mankowsky, F.~Marcellini, G.~Marinkovic, I.~Martiel,
  F.~M{\"{a}}rki, C.~J. Milne, A.~Mozzanica, K.~Nass, G.~L. Orlandi, C.~O.
  Loch, M.~Paraliev, B.~Patterson, L.~Patthey, B.~Pedrini, M.~Pedrozzi,
  C.~Pradervand, P.~Radi, J.-Y. Raguin, S.~Redford, J.~Rehanek, S.~Reiche,
  L.~Rivkin, A.~Romann, L.~Sala, M.~Sander, T.~Schietinger, T.~Schilcher,
  V.~Schlott, T.~Schmidt, M.~Seidel, M.~Stadler, L.~Stingelin, C.~Svetina,
  D.~M. Treyer, A.~Trisorio, C.~Vicario, D.~Voulot, A.~Wrulich, S.~Zerdane, and
  E.~Zimoch, \enquote{{A compact and cost-effective hard X-ray free-electron
  laser driven by a high-brightness and low-energy electron beam},}
  {\protect\JournalTitle{Nature Photonics}} \textbf{14}, 748--754 (2020).

\bibitem{Duarte2019XStereo}
J.~Duarte, R.~Cassin, J.~Huijts, B.~Iwan, F.~Fortuna, L.~Delbecq, H.~Chapman,
  M.~Fajardo, M.~Kovacev, W.~Boutu, and H.~Merdji, \enquote{{Computed stereo
  lensless X-ray imaging},} {\protect\JournalTitle{Nature Photonics}}
  \textbf{13}, 449--453 (2019).

\bibitem{Sowa2020Plenoptic}
K.~M. Sowa, M.~P. Kujda, and P.~Korecki, \enquote{Plenoptic x-ray microscopy,}
  {\protect\JournalTitle{Applied Physics Letters}} \textbf{116}, 014103 (2020).

\bibitem{Voegeli2020XMPI}
W.~Voegeli, K.~Kajiwara, H.~Kudo, T.~Shirasawa, X.~Liang, and W.~Yashiro,
  \enquote{Multibeam x-ray optical system for high-speed tomography,}
  {\protect\JournalTitle{Optica}} \textbf{7}, 514--517 (2020).

\bibitem{Goldberger20203DPtycho}
D.~Goldberger, J.~Barolak, C.~G. Durfee, and D.~E. Adams,
  \enquote{Three-dimensional single-shot ptychography,}
  {\protect\JournalTitle{Opt. Express}} \textbf{28}, 18887--18898 (2020).

\bibitem{Neutze2000SPI}
R.~Neutze, R.~Wouts, D.~van~der Spoel, E.~Weckert, and J.~Hajdu,
  \enquote{{Potential for biomolecular imaging with femtosecond X-ray pulses.}}
  {\protect\JournalTitle{Nature}} \textbf{406}, 752--757 (2000).

\bibitem{Chapman2006DiffractionBeforeDestructionExp}
H.~N. Chapman, A.~Barty, M.~J. Bogan, S.~Boutet, M.~Frank, S.~P. Hau-Riege,
  S.~Marchesini, B.~W. Woods, S.~Bajt, W.~H. Benner, R.~A. London,
  E.~Pl{\"{o}}njes, M.~Kuhlmann, R.~Treusch, S.~D{\"{u}}sterer,
  T.~Tschentscher, J.~R. Schneider, E.~Spiller, T.~M{\"{o}}ller, C.~Bostedt,
  M.~Hoener, D.~A. Shapiro, K.~O. Hodgson, D.~van~der Spoel, F.~Burmeister,
  M.~Bergh, C.~Caleman, G.~Huldt, M.~M. Seibert, F.~R. N.~C. Maia, R.~W. Lee,
  A.~Sz{\"{o}}ke, N.~Timneanu, and J.~Hajdu, \enquote{{Femtosecond diffractive
  imaging with a soft-X-ray free-electron laser},}
  {\protect\JournalTitle{Nature Physics}} \textbf{2}, 839--843 (2006).

\bibitem{Candes2006CS}
E.~J. Candès, J.~K. Romberg, and T.~Tao, \enquote{Stable signal recovery from
  incomplete and inaccurate measurements,}
  {\protect\JournalTitle{Communications on Pure and Applied Mathematics}}
  \textbf{59}, 1207--1223 (2006).

\bibitem{Donoho2006CS}
D.~Donoho, \enquote{Compressed sensing,} {\protect\JournalTitle{IEEE
  Transactions on Information Theory}} \textbf{52}, 1289--1306 (2006).

\bibitem{LeCun2015DL}
Y.~LeCun, Y.~Bengio, and G.~Hinton, \enquote{{Deep learning},}
  {\protect\JournalTitle{Nature}} \textbf{521}, 436--444 (2015).

\bibitem{kar2017learning}
A.~Kar, C.~H{\"a}ne, and J.~Malik, \enquote{Learning a multi-view stereo
  machine,} {\protect\JournalTitle{Advances in neural information processing
  systems}} \textbf{30} (2017).

\bibitem{wu2016learning}
J.~Wu, C.~Zhang, T.~Xue, B.~Freeman, and J.~Tenenbaum, \enquote{Learning a
  probabilistic latent space of object shapes via 3d generative-adversarial
  modeling,} {\protect\JournalTitle{Advances in neural information processing
  systems}} \textbf{29} (2016).

\bibitem{qi2017pointnet}
C.~R. Qi, H.~Su, K.~Mo, and L.~J. Guibas, \enquote{Pointnet: Deep learning on
  point sets for 3d classification and segmentation,} in \emph{Proceedings of
  the IEEE conference on computer vision and pattern recognition,}  (2017), pp.
  652--660.

\bibitem{chen2019learning}
Z.~Chen and H.~Zhang, \enquote{Learning implicit fields for generative shape
  modeling,} in \emph{Proceedings of the IEEE/CVF Conference on Computer Vision
  and Pattern Recognition,}  (2019), pp. 5939--5948.

\bibitem{park2019deepsdf}
J.~J. Park, P.~Florence, J.~Straub, R.~Newcombe, and S.~Lovegrove,
  \enquote{Deepsdf: Learning continuous signed distance functions for shape
  representation,} in \emph{Proceedings of the IEEE/CVF Conference on Computer
  Vision and Pattern Recognition,}  (2019), pp. 165--174.

\bibitem{mildenhall2020nerf}
B.~Mildenhall, P.~P. Srinivasan, M.~Tancik, J.~T. Barron, R.~Ramamoorthi, and
  R.~Ng, \enquote{Nerf: Representing scenes as neural radiance fields for view
  synthesis,} in \emph{European conference on computer vision,}  (Springer,
  2020), pp. 405--421.

\bibitem{yu2021pixelnerf}
A.~Yu, V.~Ye, M.~Tancik, and A.~Kanazawa, \enquote{pixelnerf: Neural radiance
  fields from one or few images,} in \emph{Proceedings of the IEEE/CVF
  Conference on Computer Vision and Pattern Recognition,}  (2021), pp.
  4578--4587.

\bibitem{henzler2021unsupervisedvideos}
P.~Henzler, J.~Reizenstein, P.~Labatut, R.~Shapovalov, T.~Ritschel, A.~Vedaldi,
  and D.~Novotny, \enquote{Unsupervised learning of 3d object categories from
  videos in the wild,} in \emph{Proc. {CVPR},}  (2021).

\bibitem{Chapman2010Coherent}
H.~N. Chapman and K.~A. Nugent, \enquote{{Coherent lensless X-ray imaging},}
  {\protect\JournalTitle{Nature Photonics}} \textbf{4}, 833--839 (2010).

\bibitem{Bravin2013PhaseContrast}
A.~Bravin, P.~Coan, and P.~Suortti, \enquote{{X-ray phase-contrast imaging:
  from pre-clinical applications towards clinics.}}
  {\protect\JournalTitle{Physics in medicine and biology}} \textbf{58}, R1--35
  (2013).

\bibitem{Andersen1984SART}
A.~Andersen and A.~Kak, \enquote{Simultaneous algebraic reconstruction
  technique (sart): A superior implementation of the art algorithm,}
  {\protect\JournalTitle{Ultrasonic Imaging}} \textbf{6}, 81--94 (1984).

\bibitem{Liu2020TomoGAN}
Z.~Liu, T.~Bicer, R.~Kettimuthu, D.~Gursoy, F.~De~Carlo, and I.~Foster,
  \enquote{Tomogan: low-dose synchrotron x-ray tomography with generative
  adversarial networks: discussion,} {\protect\JournalTitle{Journal of the
  Optical Society of America A}} \textbf{37}, 422--434 (2020).

\bibitem{Paganin2006CoherentX-ray}
D.~Paganin, \emph{Coherent X-Ray Optics} (Oxford University Press, United
  Kingdom, 2006), 1st ed.

\bibitem{Garcia-Moreno2019MetallicFoams}
F.~Garc{\'\i}a-Moreno, P.~H. Kamm, T.~R. Neu, and J.~Banhart,
  \enquote{{Time-resolved {\it in situ} tomography for the analysis of evolving
  metal-foam granulates},} {\protect\JournalTitle{Journal of Synchrotron
  Radiation}} \textbf{25}, 1505--1508 (2018).

\bibitem{Stampanoni2006TOMCAT}
M.~Stampanoni, A.~Groso, A.~Isenegger, G.~Mikuljan, Q.~Chen, A.~Bertrand,
  S.~Henein, R.~Betemps, U.~Frommherz, P.~Böhler, D.~Meister, M.~Lange, and
  R.~Abela, \enquote{{Trends in synchrotron-based tomographic imaging: the SLS
  experience},} in \emph{Developments in X-Ray Tomography V,}  vol. 6318
  U.~Bonse, ed., International Society for Optics and Photonics (SPIE, 2006),
  pp. 193 -- 206.

\bibitem{Mokso2017Gigafrost}
R.~Mokso, C.~M. Schlep{\"{u}}tz, G.~Theidel, H.~Billich, E.~Schmid, T.~Celcer,
  G.~Mikuljan, L.~Sala, F.~Marone, N.~Schlumpf, and M.~Stampanoni,
  \enquote{{GigaFRoST: the gigabit fast readout system for tomography},}
  {\protect\JournalTitle{Journal of Synchrotron Radiation}} \textbf{24},
  1250--1259 (2017).

\bibitem{xie2021neuralfield}
Y.~Xie, T.~Takikawa, S.~Saito, O.~Litany, S.~Yan, N.~Khan, F.~Tombari,
  J.~Tompkin, V.~Sitzmann, and S.~Sridhar, \enquote{Neural fields in visual
  computing and beyond,}  (2021).

\bibitem{he2016deep}
K.~He, X.~Zhang, S.~Ren, and J.~Sun, \enquote{Deep residual learning for image
  recognition,} in \emph{Proceedings of the IEEE conference on computer vision
  and pattern recognition,}  (2016), pp. 770--778.

\bibitem{rahaman2019spectral}
N.~Rahaman, A.~Baratin, D.~Arpit, F.~Draxler, M.~Lin, F.~Hamprecht, Y.~Bengio,
  and A.~Courville, \enquote{On the spectral bias of neural networks,} in
  \emph{International Conference on Machine Learning,}  (PMLR, 2019), pp.
  5301--5310.

\bibitem{tancik2020fourier}
M.~Tancik, P.~P. Srinivasan, B.~Mildenhall, S.~Fridovich-Keil, N.~Raghavan,
  U.~Singhal, R.~Ramamoorthi, J.~T. Barron, and R.~Ng, \enquote{Fourier
  features let networks learn high frequency functions in low dimensional
  domains,} {\protect\JournalTitle{arXiv preprint arXiv:2006.10739}}  (2020).

\bibitem{vaswani2017attention}
A.~Vaswani, N.~Shazeer, N.~Parmar, J.~Uszkoreit, L.~Jones, A.~N. Gomez,
  {\L}.~Kaiser, and I.~Polosukhin, \enquote{Attention is all you need,} in
  \emph{Advances in neural information processing systems,}  (2017), pp.
  5998--6008.

\bibitem{kingma2014adam}
D.~P. Kingma and J.~Ba, \enquote{Adam: A method for stochastic optimization,}
  {\protect\JournalTitle{arXiv preprint arXiv:1412.6980}}  (2014).

\bibitem{ronneberger2015u}
O.~Ronneberger, P.~Fischer, and T.~Brox, \enquote{U-net: Convolutional networks
  for biomedical image segmentation,} in \emph{International Conference on
  Medical image computing and computer-assisted intervention,}  (Springer,
  2015), pp. 234--241.

\end{thebibliography}


\begin{thebibliography}{1}
\newcommand{\enquote}[1]{``#1''}

\bibitem{Andersen1984SART}
A.~Andersen and A.~Kak, \enquote{Simultaneous algebraic reconstruction
  technique (sart): A superior implementation of the art algorithm,}
  {\protect\JournalTitle{Ultrasonic Imaging}} \textbf{6}, 81--94 (1984).

\bibitem{Garcia-Moreno2019MetallicFoams}
F.~Garc{\'\i}a-Moreno, P.~H. Kamm, T.~R. Neu, and J.~Banhart,
  \enquote{{Time-resolved {\it in situ} tomography for the analysis of evolving
  metal-foam granulates},} {\protect\JournalTitle{Journal of Synchrotron
  Radiation}} \textbf{25}, 1505--1508 (2018).

\bibitem{Stampanoni2006TOMCAT}
M.~Stampanoni, A.~Groso, A.~Isenegger, G.~Mikuljan, Q.~Chen, A.~Bertrand,
  S.~Henein, R.~Betemps, U.~Frommherz, P.~Böhler, D.~Meister, M.~Lange, and
  R.~Abela, \enquote{{Trends in synchrotron-based tomographic imaging: the SLS
  experience},} in \emph{Developments in X-Ray Tomography V,}  vol. 6318
  U.~Bonse, ed., International Society for Optics and Photonics (SPIE, 2006),
  pp. 193 -- 206.

\end{thebibliography}

\end{document}


\title{ONIX: an X-ray deep-learning tool for 3D reconstructions from sparse views: supplemental document}
\author{} 

\begin{abstract}
This document provides supplementary information to “ONIX: an X-ray deep-learning tool for 3D reconstructions from sparse views.” 
In this material, we report supplementary figures of the results of \name for both simulation and experimental datasets.
\end{abstract}

\setboolean{displaycopyright}{false} 

\maketitle

\section{Supplementary figures}

This section presents two figures with slices over the 3D volumes used and retrieved from the simulation and experimental dataset.
Finally, we present a study over the \name reconstructions obtained by different combinations of four and six encoded views using the experimental dataset.

\begin{figure}[htbp!]
\centering
\includegraphics[width=0.8 \linewidth]{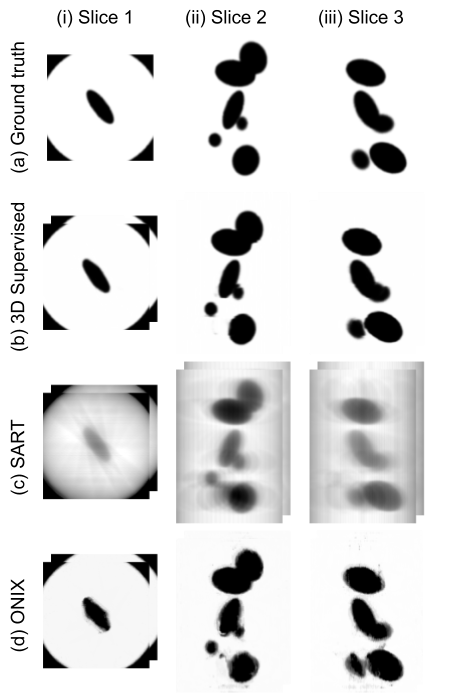}
\caption{
Reconstructed slices from different methods for the simulation dataset. (a) The simulated ground truth. (b-d) Reconstructed results from (b)3D supervised learning, (c) \ac{SART}, and (d) \name. 
The three columns show the three slices at the center of the three orthogonal axes, denoted by (i)slice1, (ii)slice2, and (iii)slice3.
We plotted only the absorption images overlaid with the phase for the reconstructions of the three methods.}
\label{fig:ellipsoid-slice}
\end{figure}

First, we present the results of \name applying to the simulation dataset. 
The \ac{DL} methods were trained using 1000 simulated 3D objects, where we simulated randomly placed ellipsoids inside a cylinder.
We compared \name results to the ground truth, the 3D supervised learning results, and the \ac{SART}~\cite{Andersen1984SART} results.
The results are shown in Fig.~\ref{fig:ellipsoid-slice}, where we plotted the three slices at the center of the three orthogonal axes, denoted by (i) slice 1, (ii) slice 2, and (iii) slice 3, for the (a) ground truth, and the results from (b) 3D supervised learning, (c) SART, and (d) \name.
From Fig.~\ref{fig:ellipsoid-slice} we can see that the best reconstructions were achieved by the 3D supervised learning approach. 
Nonetheless, 3D supervised learning requires the 3D information (ground truth) of the training data, limiting its application.
On the other hand, it can be seen from the figure that, \name, outperformed \ac{SART} method and provided decent reconstructions with no 3D supervision.

\begin{figure}[htbp!]
\centering
\includegraphics[width=1 \linewidth]{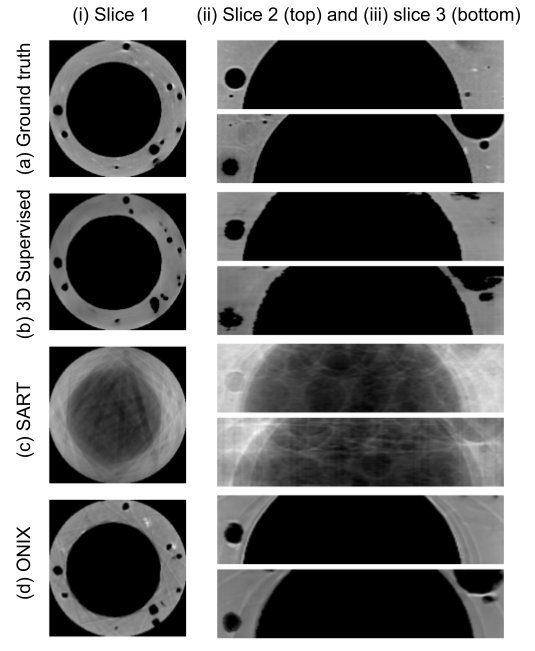}
\caption{Reconstructed slices from different methods. (a) Ground truth from \ac{SART} using 96 projections. (b-d) Reconstructed results from (b)3D supervised learning, (c) \ac{SART}, and (d) \name. 
The three slices (i-iii) at the center of the three orthogonal axes are plotted for the ground truth and the reconstructed results. The size between slices was adjusted for better arrangement.}
\label{fig:thixo-slice}
\end{figure}

Next, we show the reconstruction results of \name applying on the experimental metallic foam dataset~\cite{Garcia-Moreno2019MetallicFoams} collected from the TOMCAT beamline at the Swiss Light Source~\cite{Stampanoni2006TOMCAT}.
Eight projections were used in the training of the \ac{DL} methods and also for the \ac{SART} Reconstruction. 
We used the reconstructions from \ac{SART} using 96 projections as the ground truth.
Fig.~\ref{fig:thixo-slice} shows the (a) ground truth and the reconstructed results from (b) 3D supervised learning, (c) SART, and (d) \name. We plotted the three center slices of the three orthogonal axes (i-iii), the same as for the simulation dataset.
It can be seen from the figure that the experimental dataset was more complicated and contained more details compared to the simulation dataset.
The reconstructions from both \ac{DL} methods outperformed the \ac{SART} results, providing clear features in the slices. 
The 3D supervised results, as trained on slices, failed to capture the circular shape of the metallic foam in the direction perpendicular to the training slices, as shown in Fig.~\ref{fig:thixo-slice} b (iii).
\name, on the other hand, learned better the structure of the foam even without accessing the 3D information provided by the ground truth.

\begin{figure}[htbp!]
\centering
\includegraphics[width=1 \linewidth]{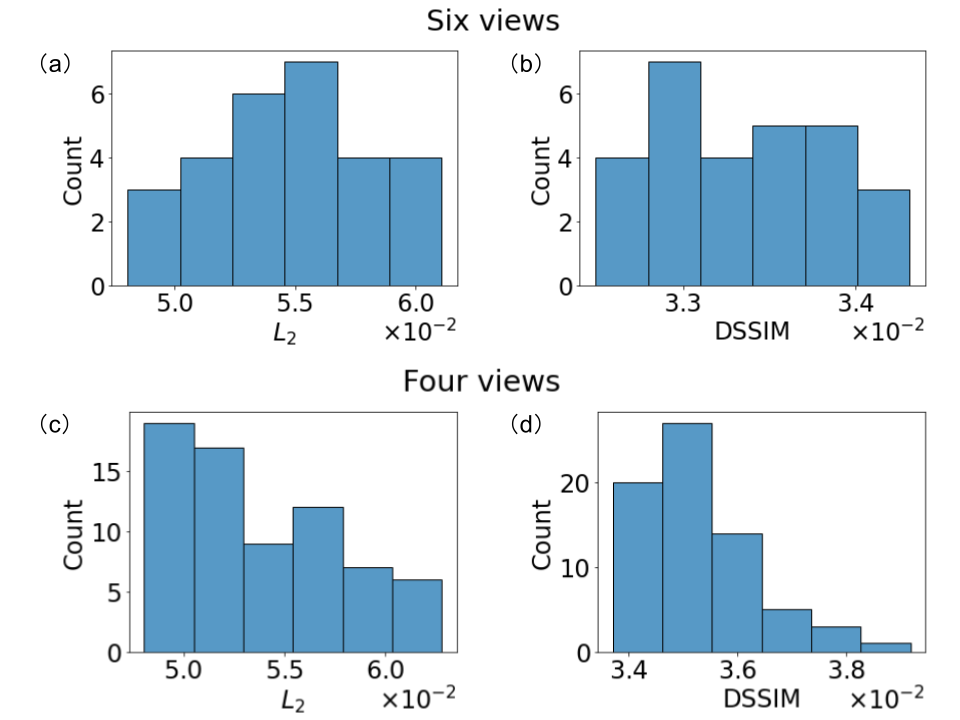}
\caption{Distribution of \name results reconstructed using all possible combinations of four and six views over eight measured projections. (a,c) The distribution of $L_2$ norm for the six-view and four-view results. (b, d) The distribution of \ac{DSSIM} for the six-view and four-view results.}
\label{fig:histogram}
\end{figure}

Finally, we show the dependency of \name on using different encoded views to retrieve the 3D volume from eight views studying the experimental dataset. 
The eight projections used to train \name for each tomographic experiment were equally spaced between $0^\circ$ to $131^\circ$.
Although eight views were used for training, only a subset of them was used for the reconstructions to constrain the retrieved 3D volume at a time.
Specifically, we evaluated the quality of the reconstructions using four and six encoded views to constraint the 3D reconstructions, and the results are summarized in Fig.~\ref{fig:histogram}.
The \name reconstructions were compared to the ground truth via two evaluation metrics: $L_2$ norm and \ac{DSSIM}.
Fig.~\ref{fig:histogram} shows the distribution of the reconstructions results from all the different combinations of the encoded views.  
Note that the six-view and four-view results came from two different trained \name models, each of them trained using six or four random encoded views in each training iteration. 
The different view combinations were only used in the reconstruction stage, where we applied different encoded views to the trained model to generate 3D reconstructions. 

It can be seen that \name is capable of performing tomographic reconstruction for all of the experiments we did.
For the reconstructions of four encoded views, the values of the $L_2$ norm are similar to the one from six views, but the deviation of \ac{DSSIM} is more significant for the four-view case.
We see that by adding more encoded views in training, the distribution of the reconstructions shifted towards a Gaussian distribution, implying more reliable reconstructions. 

\bibliography{Biblio}